\documentclass{article}
\usepackage{arxiv}

\usepackage{url}            
\usepackage{booktabs}       
\usepackage{amsfonts}       
\usepackage{nicefrac}       
\usepackage{microtype}      
\usepackage{authblk}        
\usepackage{amsmath}
\usepackage{gensymb}
\usepackage{graphicx}
\usepackage{placeins}

\newcommand{\UCParticle}{$\beta$-NaYF\textsubscript{4}:(18\%Yb\textsuperscript{3+}, 3\%Er\textsuperscript{3+})}
\newcommand{\nm}{$\;\textrm{nm}$}
\title{Double-Layer Metasurface for Enhanced Photon Up-Conversion}
\author[1,2,$\ddag$]{Phillip Manley*}
\author[1,$\ddag$]{Michele Segantini}
\author[1]{Doguscan Ahiboz}
\author[3]{Martin Hammerschmidt}
\author[4]{Georgios Arnaoutakis}
\author[1]{Rowan W.MacQueen}
\author[2,3]{Sven Burger}
\author[1]{Christiane Becker*}
\affil[1]{Helmholtz Zentrum Berlin f{\"u}r Materialien und Energie, Berlin, Germany}
\affil[2]{Zuse Institute Berlin, Berlin, Germany}
\affil[3]{JCMwave, Berlin, Germany}
\affil[4]{Ben-Gurion University, Israel; current address: Hellenic Mediterranean University, Greece}
\affil[$\ddag$]{These authors contributed equally to this work.}
\affil[*]{Corresponding authors: phillip.manley@helmholtz-berlin.de, christiane.becker@helmholtz-berlin.de}
\begin{document}
\maketitle

\begin{abstract}
	We present a double-layer dielectric metasurface obtained by stacking a silicon nanodisc array and a silicon photonic crystal slab with equal periodicity on top of each other. We focus on the investigation of electric near-field enhancement effects occurring at resonant excitation of the metasurface and study its optical properties numerically and experimentally. We find that the major difference in multi-layer metasurfaces when compared to conventional single-layer structures appears to be in Rayleigh-Wood anomalies: they are split into multiple different modes which are themselves spectrally broadened. As a proof of concept we cover a double-layer metasurface with a lanthanide-doped up-conversion particle layer and study its interaction with a 1550 nm photoexcitation. We observe a 2.7-fold enhancemed up-conversion photoluminescence by using the stacked metasurface instead of a planar substrate, although only around 1\% of the up-conversion material is exposed to enhanced near-fields. Two mechanisms are identified explaining this behavior: First, enhanced near-fields when exciting the metasurface resonantly, and second, light trapping by total internal reflection in the particle layer when the metasurface redirects light into high-angle diffraction orders. These results pave the way for low-threshold and, in particular, broadband photon up-conversion in future solar energy and biosensing applications.

\end{abstract}


\section{Introduction}
\label{sec:Introduction}

Metasurfaces have recently received great interest for their ability to control the amplitude, phase, polarization state and angular distribution of light, all using structures with sub-wavelength thickness \cite{Chen2016}. This unlocks the potential for ultra-thin applications including phased arrays \cite{Sun2013}, achromatic correctors \cite{Chen2018}, beam splitters \cite{Zhang2018}, chiral holograms \cite{Mueller2017}, and anomalous reflection \cite{Chen2018}. The working principle behind all of these effects is the periodic arrangement of individual resonant structures. 
The collective behaviour of the structures gives rise to new resonances, which may be entirely different to those of the individual structures. 
Thus, a metasurface has degrees of freedom related to the composition and geometry of the individual resonators and to their arrangement in the 2D plane. Further, metasurfaces are much more practical for standard fabrication methods such as lithography when compared to fully 3D metamaterials \cite{Becker2014}.

A further opportunity for tuning metasurface resonances comes from using multiple layers of scattering structures to further enhance the manipulation of light. The layers may be spatially separated, in which case they are coupled together via far field interactions. This has been applied to the concept of retroreflection \cite{Arbabi2017} and to light bending \cite{Pfeiffer2014}. Multi-layer metasurfaces have been used for spectral splitting, by creating a stacked metasurface with different layers interacting with different wavelengths of light \cite{Ollanik2018}, this avoids coupling between the different layers, as they have little spectral overlap. In general, such far field coupled multi-layer metasurfaces can be described as multiple, distinct metasurfaces, the net effect of which can be determined via matrix manipulations, should the response of each metasurface be known \cite{Menzel2016}. 

More complex behavior may arise by allowing for near field interactions between different layers of the metasurface. By combining multiple layers of scattering structures, extremely compact and high efficiency lenses have been fabricated \cite{Monticone2013}. Furthermore, the two concepts of 2D materials and metasurfaces have been combined in a stacked graphene metasurface which was able to tune the phase of reflected light \cite{Ma2017}.

Many applications of metasurfaces have also focused on their ability to manipulate light fields. The resonant behavior which produces these manipulations is associated with strongly enhanced near fields. These near-fields can be exploited for the enhancement of non-linear processes in secondary materials which find themselves within the near fields of the metasurface. This idea has been used in the detection of mid-infrared spectral fingerprints for the composition analysis of surface-bound analytes \cite{Tittl2018}, as well as for enhanced second-harmonic generation \cite{Kruk2015}.

One area in which metasurface induced near field enhancement has found particular success is photon up-conversion. Photon up-conversion materials, such as lanthanide-doped particles, exploit a variety of energy-transfer mechanisms to convert multiple low-energy photoexcited states into a single higher-energy state, which can re-emit the pooled energy as a shorter wavelength photon, producing the photon up-conversion effect \cite{Guo2018}. The efficiency of this up-conversion process can be greatly enhanced by increasing the photoexcitation rate of the system, i.e. the light intensity. In order to obtain high up-conversion efficiency at low incident intensity, near fields can locally increase the incident power for lanthanide doped nanoparticles close to a metasurface \cite{Wuerth2020, Mao2019}. This has been accomplished using metasurfaces consisting of 2D anodic aluminium oxide photonic crystals \cite{Wang2016}, SiO\textsubscript{2} opal photonic crystals \cite{Shi2019}, and nanopillar based photonic crystals \cite{Gong2019}.

An alternative to metasurface-based enhancement of the up-conversion process is superlensing via microparticles \cite{Liang2019,Liu2019}. The downside to superlensing is the requirement of careful alignment of the up-conversion material with the focal point of the lense. However, depending on the application, this may not pose a significant drawback.

Most of the results for the enhancement of the up-conversion process refer to an excitation wavelength of 980\nm. This wavelength can be used for the efficient generation of visible photons, which makes it attractive for many applications, such as the detection of biological agents by eye \cite{Zhang2019}. Up-conversion using lanthanide-based nanoparticles can also support transitions excited at a wavelength of 1550\nm, resulting in emission, primarily, at 980\nm. This can be used for increasing the efficiency of photovoltaic devices, by converting below band-gap photons to a wavelength that can be absorbed by silicon \cite{Goldschmidt2015}. There has already been a report of up-conversion enhancement at 1550\nm\ using metallic nanostructures \cite{Christiansen2020}. All these experiments have in common that single-layer metasurfaces are applied. The effect of stacking metasurfaces on their ability to provide enhanced near-fields has, to the best of our knowledge, not been investigated so far.

This work presents a large-area, stacked, double-layered silicon metasurface based on nanoimprint lithography. The metasurface is analyzed with regard to near-field enhancement effects. Transmission measurements are used to identify the spectral positions of resonances, with simulations confirming that the resonances are associated with enhanced near fields. In order to understand the provenance of the resonant modes, each layer of the metasurface is simulated separately and their contribution to the complete metasurface is discussed. As a proof-of-concept the double-layer metasurfaces is covered with lanthanide doped up-conversion particles and their interaction with resonances of the metasurface is analyzed by measuring up-conversion fluorescence upon excitation at 1550\nm. This results in increased up-conversion efficiency, which we ascribe to a combination of enhanced near fields of the metasurface and light trapping effects due to high angle scattering. Finally, we present a 4-layer stacked metasurface and discuss its potential for photon up-conversion applications upon low-intensity and broadband excitation.

\section{Method}
\label{sec:Method}

The silicon metasurface was fabricated using a method based on nanoimprint lithography (NIL), physical evaporation and thermal crystallization of silicon. The starting point was a $5\times 5\;\textrm{cm}$\textsuperscript{2} master structure provided by a commercial manufacturer (EULITHA, Switzerland). The master exhibits a hexagonal array of nanopillars with lattice constant $1000$\nm, a width of $400$\nm\ and a height of $500$\nm. A stamp was created by molding polydimethylsiloxane (PDMS) (Wacker, Germany) onto the master structure, which was subsequently cured at 80${\degree}$C for 3 hours. This stamp was imprinted on a glass substrate covered by a UV-thermally curable sol-gel layer (Philips, Netherlands), which was hardened under UV illumination for 500\,s. Immediately after, a thermal curing at 100$\degree$C for 8 minutes was applied in order to evaporate any residual solvent on the surface for the sample. Subsequently, a post-deposition thermal curing at 600$\degree$C was performed for 1 hour to shrink the structures by approximately 10\% compared to the initial master structure, however, the pitch of 1000\,nm is not affected. This thermal curing causes surface deformations to form, which aid the subsequent silicon deposition. In the final step, the sample was covered with a layer of amorphous silicon (a-Si) by physical vapor deposition and thermally crystallized at 600$\degree$C under nitrogen flow to form crystalline silicon (c-Si). The resulting silicon layer on glass has a thickness of around $85$\nm. A double-layer metasurface is formed consisting of round silicon discs, also with around $85$\nm\ thickness, on top of the SiO$\mathrm{_x}$ pillars with around 400\nm\ thickness. The diameter of both tips and pillars is around 410\nm. The base of each SiO$\mathrm{_x}$ pillar is in one of the periodically arranged holes in the silicon layer. An optical image of the metasurface is presented in figure \ref{fig:SamplePictures}(a), left side. Parts (b,c) show scanning electron microscopy images of the metasurface at different magnifications. 

In order to fabricate an upconverting layer, lanthanide-doped \UCParticle\ crystals (Leuchtstoffwerk Breitungen, Germany) were dispersed in water. An aqueous solution of ethylene diamine tetraacetic acid (EDTA) at 1:10 microcrystal:EDTA ratio was dropwise added to the emulsion. The mixture was stirred for 1 hour and finally dried in air for 12 hours. The dried \UCParticle\ microcrystals were dispersed in deionized water in a glass vial, followed by immersion of the glass vial in a sonic bath. The silicon metasurface ($1.1\;\textrm{cm} \times 2.2\; \textrm{cm}$) was covered by a thin up-converting layer. First, $0.096\;\textrm{g}$ of particulate matter was dispersed in $100\;\mu\textrm{L}$ of toluene by sonication. This solution was mixed with $0.115\;\textrm{g}$ of PDMS with a magnetic stirrer for 1 hour. Then, the metasurfce was covered by spin-coating at two different speeds: $500\;\textrm{rpm}$ for $12\;\textrm{s}$ and $3000\;\textrm{rpm}$ for $20\;\textrm{s}$. Finally, the sample was baked in a vacuum oven at 60$\degree$C for 1 hour. The final result is shown in figure \ref{fig:SamplePictures}(a), right side. The upconverting layer obtained after the fabrication process was measured by a profilometer at various positions, with a resulting homogeneous thickness of $15\;\mu\textrm{m}$, as shown in figure \ref{fig:SamplePictures}(b).

The angular-resolved directional transmittance (ARDT) measurements were performed using a PerkinElmer Lambda 1050 Spectrometer with an ARTA (OMT solution) tool. The sample was rotated along the $\Gamma-\mathrm{K}$ high symmetry direction using TM polarized light.

The optical simulations were performed using the commercial finite element (FE) solver JCMsuite \cite{Pomplun2007}. The time harmonic Maxwell equations were solved with an exterior plane wave source. A typical mesh used for simulation is shown in figure \ref{fig:SamplePictures}(d). Bloch periodic boundary conditions were used for boundaries in the $x-y$ plane, while transparent (perfectly matched layer) boundary conditions were used in the $z$ direction. Reflections at the planar interface at the rear of the glass substrate and the top of the dispersed nanoparticle coating were neglected. The mesh size was adapted to be at least as small as one quarter of the wavelength inside each material. The transmittance and reflectance were determined using the Fourier transform of light leaving the periodic domain. The local electric field energy enhancement was calculated by integrating the electric field energy in a volume of space between and above the nanopillars of the metasurface up to 200\nm\ from the tip surface. This volume is labeled with coating in figure \ref{fig:SamplePictures}(d). This value was then normalised to the electric field energy in the same volume of material for a plane wave. This was directly used as a measure for the enhanced fluorescence from the up-conversion particles. The refractive index of the glass substrate \cite{Cushman2017}, silicon \cite{Palik1998}, SiOx \cite{Gao2013}, and lanthanide particles in solvent \cite{Sokolov2015} were taken from literature data.

Fluorescence emission measurements of the lanthanide-doped particles prior to deposition were performed using a Thorlabs CCS200 spectrometer with a detector range of 200-1020\nm. For the angular resolved fluorescence emission measurements each sample was placed inside an integrating sphere. The sample was then rotated along the $\Gamma$-K direction and illuminated by TM polarized light. The intensity of the laser light was measured in front of the integrating sphere and kept constant at 3.7Wcm\textsuperscript{-2}. These measurements were performed on the coated metasurface and a coated planar Si sample. 

In order to characterise a nonlinear optical process, such as up-conversion fluorescence, the dependence of the signal on the excitation intensity can be analyzed. The peak emission signal will be a function of the excitation intensity with a certain exponent. For linear processes the exponent would be one, for higher values the process is said to be non-linear. By taking the logarithm of both the fluorescence signal and excitation intensity a linear fit is able to determine the exponent \cite{Fischer2015},
\begin{equation}
\label{eq:linear_fit}
\log(S)=m\;\log(I)+c\;
\end{equation}
where $\textit{S}$ is the measured emission signal (in a.u.), and $\textit{I}$ is the intensity of the laser pump. The gradient of the linear fit, $\textit{m}$, can be used to estimate the number of photons involved in the emission process.
\section{Results and Discussion}
\label{sec:ResultsAndDiscussion}
\subsection{Double-Layer Metasurface Characterization}
\begin{figure*}[h!]
    \centering
    \includegraphics[width=1.0\textwidth]{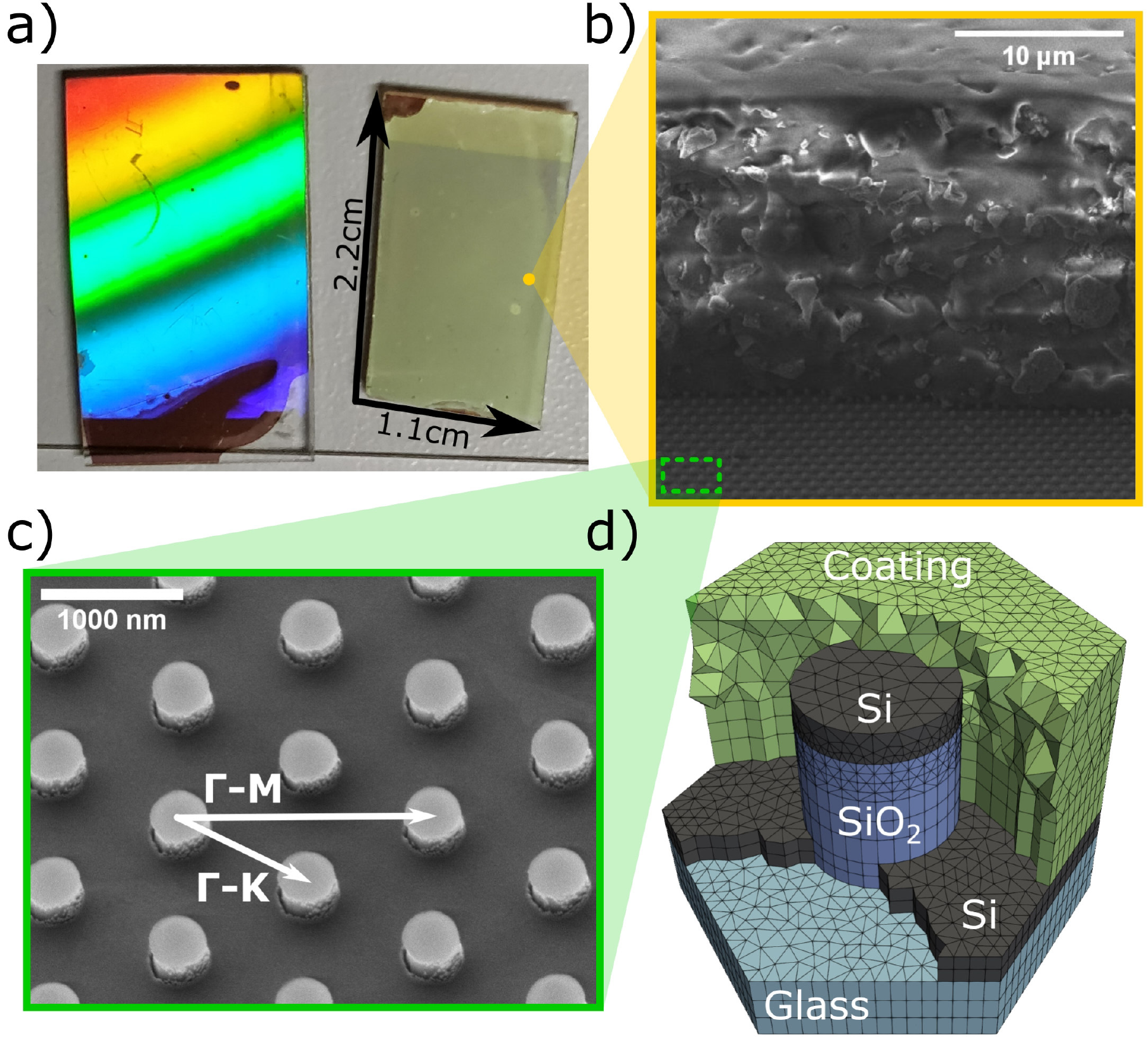}
    \caption{a) Photograph of the uncoated double metasurface (left) and the coated metasurface (right). b) Side view SEM image of the coated metasurface with an uncovered part of the metasurface at the front. c) SEM image of the uncoated periodic double-layer metasurface with lattice vectors. d) Mesh of the simulation unit cell with layers partially removed for clarity.
    }
    \label{fig:SamplePictures}
\end{figure*}

The results of the large area deposition of the metasurface can be seen in figure \ref{fig:SamplePictures}. The dispersion of visible light into constituent colors present in the left hand sample in figure \ref{fig:SamplePictures}(a) indicates the presence of geometrical features with a size on the same order of magnitude as the wavelength. In this case, the iridescence is caused by the uncoated double layer metasurface. The sample on the right hand side has been coated with the layer of PDMS with embedded UC particles. Due to Mie scattering of the particles, the iridescence caused by the metasurface is disrupted causing the film to appear a hazy whitish color. The upper part of the sample, which displays a lighter color, has an unstructured silicon film under the PDMS, while the visibly darker region below has the double-layer metasurface under the PDMS. At optical wavelengths the Si metasurface will be absorptive which reduces the light intensity, in turn making the sample appear darker.

Figure \ref{fig:SamplePictures}(b) presents a SEM image of the metasurface partially coated by the PDMS layer with embedded up-conversion particles at a 40$\degree$ viewing angle. The thickness of the PDMS layer containing these particles is around $15~{\mu}\textrm{m}$. In contrast, the thickness of the stacked metasurface, described in section \ref{sec:Method}, is only around $0.5\,\mu$m. Because of this extreme size mismatch and the fact that enhanced near-fields are expected within only a few hundreds of nanometers of the metasurface, the ability of the double-layer metasurface to provide an enhancement to the up-conversion process has to be evaluated. The metasurface is shown in more detail in figure \ref{fig:SamplePictures}(c), where the hexagonal lattice structure is clearly visible. The two independent lattice vectors are indicated, as well as the high symmetry directions along which the vectors lie.

The simulation mesh of the hexagonal unit cell is shown in figure \ref{fig:SamplePictures}(d) with the silicon and coating layers partially obscured for clarity. The volume of green tetrahedra labeled with "coating" serves as both the suspended particle coating and also air in the simulations of the uncoated metasurface. This is the volume used for determining the average electric field energy enhancement.

\begin{figure*}[h!]
    \centering
    \includegraphics[width=1.0\textwidth]{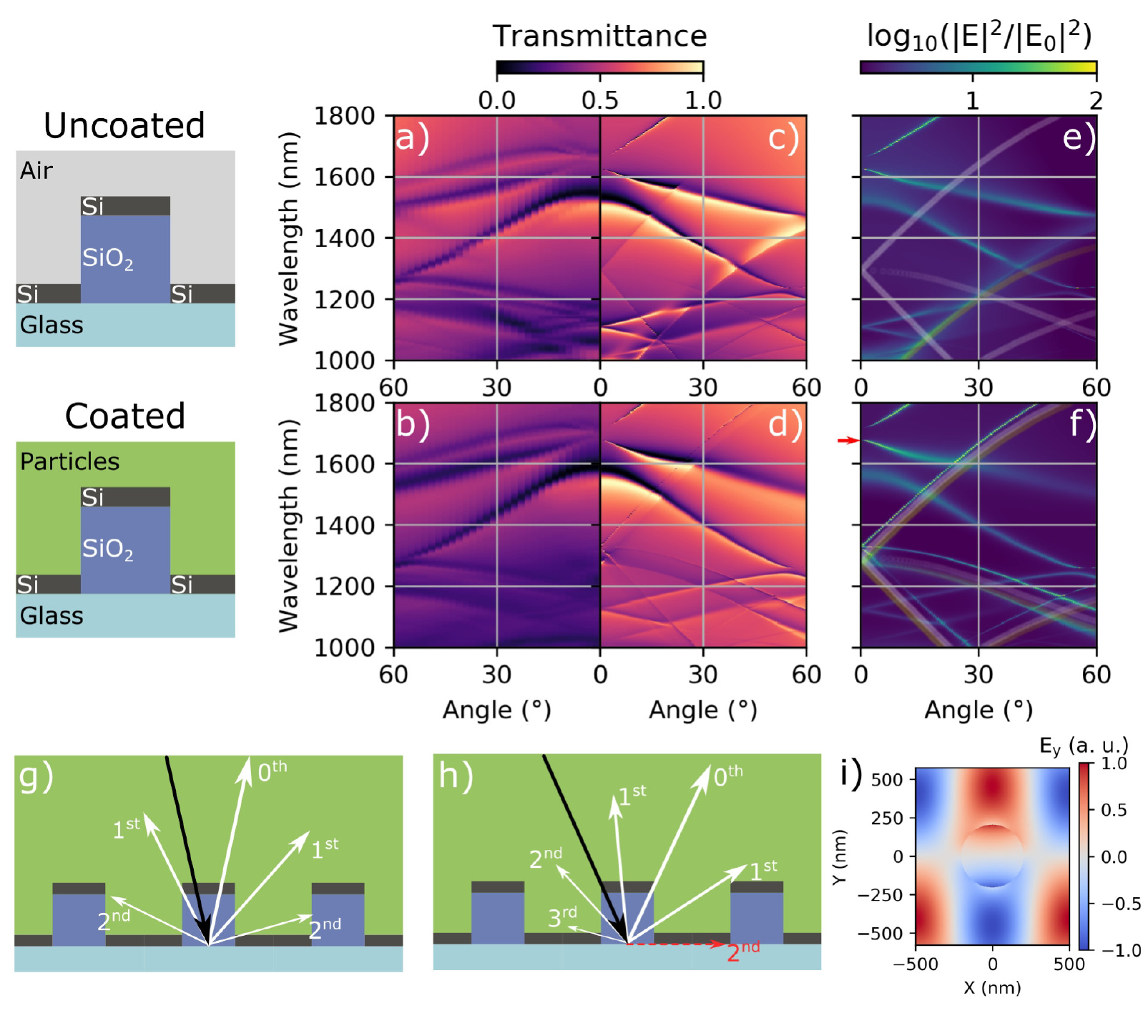}
    \caption{ARDT measurements of the uncoated (a) and coated (b) double-layer metasurface. Simulated transmittance of the uncoated (c) and coated (d) double-layer metasurface. Logarithm of the simulated electric field intensity enhancement for uncoated (e) and coated (f) double-layer metasurface. The spectral positions at which the reflection (yellow) and transmission (white) diffraction orders disappear is shown as an overlay. Example of reflectance diffraction orders of a periodic structure such as the double-layer metasurface (g). In (h) the incident angle of light has increased, causing one of the second order diffractions to become evanescent. (i) The real part of the $y$ component of the electric field in the $x-y$ plane situated at the center of the lower silicon layer of the double-layer metasurface. The field corresponds to a resonant mode solution at 1676\nm\ wavelength, indicated by the red arrow in (f).
    }
    \label{fig:ARDT}
\end{figure*}

Figure \ref{fig:ARDT}(a,b) shows angular resolved directional transmittance measurements for both the uncoated and coated metasurface using TM polarized light with rotation along the $\Gamma-\mathrm{K}$ axis. The simulated total transmittance (c,d) is also presented for comparison. The spectrum of leaky mode resonances present in the metasurface is easily identified by the wavelength and angle combinations that lead to rapid changes in the transmittance. In general excellent agreement is observed between experiment and simulation. A set of narrowband resonances appear in the simulations which are mainly absent in the experiment, except for angles close to 60$\degree$ and wavelengths between 1400 and 1500\nm. These resonances will be discussed in more detail later. The other notable difference between simulation and experiment is for wavelengths above 1600\nm. Here, a resonance with a relatively flat dispersion (wavelength decreases only slowly with increasing angle) is seen in the experimental side, while the mode with a similar normal incident spectral position presents a high negative dispersion (wavelength increases rapidly with increasing angle).

The leaky modes are associated with enhanced electric fields close to the surface. The simulated electric field energy in the volume directly above the metasurface is shown in figure \ref{fig:ARDT}(e,f) on a logarithmic scale. The presence of the same pattern of modes confirms that the transmission resonances are associated with enhanced near fields. Since photon up-conversion scales non-linearly with the light intensity, these enhanced near fields can efficiently drive up-conversion.

Also plotted in figure \ref{fig:ARDT}(e,f) are the critical angles of incidence and their associated wavelengths at which the angle of a diffraction order becomes 90$\degree$ with respect to the normal. Figure \ref{fig:ARDT}(g,h) presents schematic explanations of the process. At this critical angle the diffraction order transitions from being a radiating mode to an evanescent one. This is often associated with a sharp change in the reflection and transmission of a diffractive structure, often referred to as a Rayleigh-Wood anomaly \cite{Wood1902,Rayleigh1907,Garcia2007}. Due to the fact that these critical angles depend only on the geometry of the periodic repeating structure and the refractive index of the half space into which the diffraction orders radiate, the critical angles and their wavelength dependence can be calculated analytically. For each of these Rayleigh-Wood anomalies a corresponding resonance mode of the metasurface can be identified both in the simulation of enhanced electric field energy and transmission. Interestingly, these modes are the ones discussed above which are absent from the transmittance measurements, apart from the 1st order diffraction mode in air, which is faintly visible in the region between 30$\degree$ and 60$\degree$ and 1400 to 1500\nm\ wavelength in figure \ref{fig:ARDT}(a). We note also that this resonance is relatively broad compared to the other Rayleigh-Wood anomalies. The other Rayleigh-Wood anomalies are likely absent from the transmission measurements due to two factors. They are resonances with a very narrow spectral width, meaning that the illumination source with a spectral width of 2 nm is not able to fully excite the resonance. Related to the fact that the resonances are so narrow, their quality factors are also very high. High quality factor resonances are more susceptible to being quenched by imperfections in the sample geometry. The resonant field shape associated with these modes (figure \ref{fig:EnhancementAnalysis}) also reveals modes that appear plane wave like with propagation in the $x-y$ plane and polarisation in the $z$ axis, which corresponds to a disappearing diffraction order. 

Of the remaining resonances, one belongs to a type of resonance which has been the subject of intense interest in the recent literature \cite{Hsu2016, Stillinger1975}. The topic of bound in continuum (BIC) states has spurred interest not only due to their fundamental nature but also due to their ability to create extremely high quality factor resonances. BIC resonances can be categorised into two classes, symmetry protected and incidental. The resonance  mode present in figure \ref{fig:ARDT}, visible for all angles and for wavelengths between 1500 and 1600~nm, belongs to the symmetry protected class \cite{Cong2019}. This can be seen due to the fact that the resonance disappears at the $\Gamma$ point, where the symmetry of the resonant mode is incompatible with that of the zeroth order diffraction (i.e. a plane wave propagating normal to the surface). Furthermore, the farther from the $\Gamma$ point, the wider and less intense the resonance becomes. This is typical behaviour for a symmetry protected BIC as reported in the literature \cite{Li2019}. The field distribution of the resonant mode found at a wavelength of 1676\nm\ (spectral position indicated with the red arrow in figure \ref{fig:ARDT}(f)) is shown in figure \ref{fig:ARDT}(i). The symmetry of the mode is odd with respect to a rotation of $\pi$ around the $z$ axis. The field of a plane wave is even with respect to such a rotation, thus preventing coupling to the resonance. 

All other resonances visible in the transmittance spectrum we attribute to coupling to leaky eigenmodes of the photonic crystal lattice. This means that they are affected not only by the periodicity (as with the Rayleigh-Wood anomalies), or the symmetry of the unit cell (as for the BIC) but also geometrical quantities like the PhC hole diameter. This allows for greater flexibility in tailoring such resonances for the enhancement of non-linear processes such as up-conversion. 

\begin{figure*}[h!]
    \centering
    \includegraphics[width=1.0\textwidth]{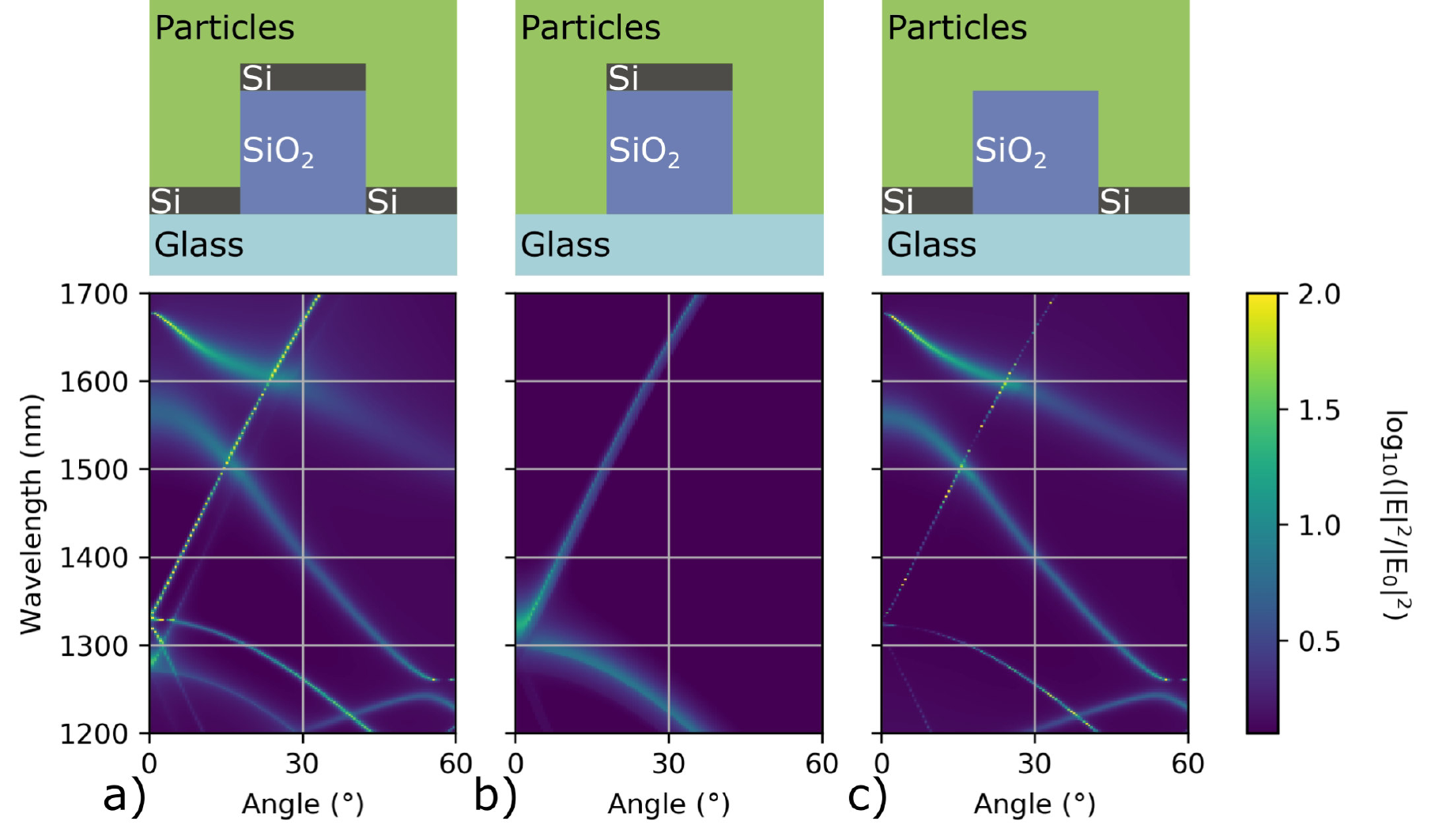}
    \caption{The logarithm of the electric field intensity enhancement in the coating region directly above the metasurface. Three different geometries are considered: the double-layer metasurface (a), the silicon tips only (b) and the silicon photonic crystal only (c). Schematic cross section images of the geometry are shown above.}
    \label{fig:GeometryVariation}
\end{figure*}


In order to understand the mechanism of the resonant modes in more detail, we simulate the constituent parts of the double-layer metasurface in isolation. In addition to the coated double-layer metasurface, two further simulations were performed with either the silicon tips or the lower silicon photonic crystal slab removed. Figure \ref{fig:GeometryVariation} shows the electric field intensity enhancement in the volume above the surface for each of the three cases. The simulations of the metasurface without the silicon tips (c) retains largely the same resonance structure as for the double-layer metasurface (a), with the resonances being somewhat narrower, especially around normal incidence. The largest difference is with the sharp Rayleigh-Wood anomaly type resonances. The double-layer metasurface clearly has two sets of these resonances, whereas the photonic crystal slab metasurface exhibits only one set. When moving to the metasurface consisting of the silicon tips (b), only the Rayleigh-Wood anomalies are present, and they are significantly broadened compared to the for the other two metasurfaces. It should be noted that the Rayleigh-Wood resonances for the double-layer metasurface are spectrally shifted compared to the two cases of the individual layers simulated in isolation. Due to both layers of the metasurface having the same period, the Rayleigh-Wood anomalies will spectrally overlap. For the double layer metasurface, these resonances interact in such a way as to separate them spectrally.

To summarize, the major difference in the double-layer metasurface when compared to a conventional single-layer photonic crystal slab structure appears to be in the Rayleigh-Wood anomalies, which are split into two different modes and spectrally broadened. This can potentially be useful for applications with excitation sources with significant spectral bandwidths. As a proof of concept for near field enhancement capabilities of the double-layer metasurface we now move to fluorescence measurements of the lanthanide particles in PDMS deposited directly onto the metasurface.

\subsection{Up-conversion Measurements}
\begin{figure*}[ht!]
    \centering
    \includegraphics[width=0.8\textwidth]{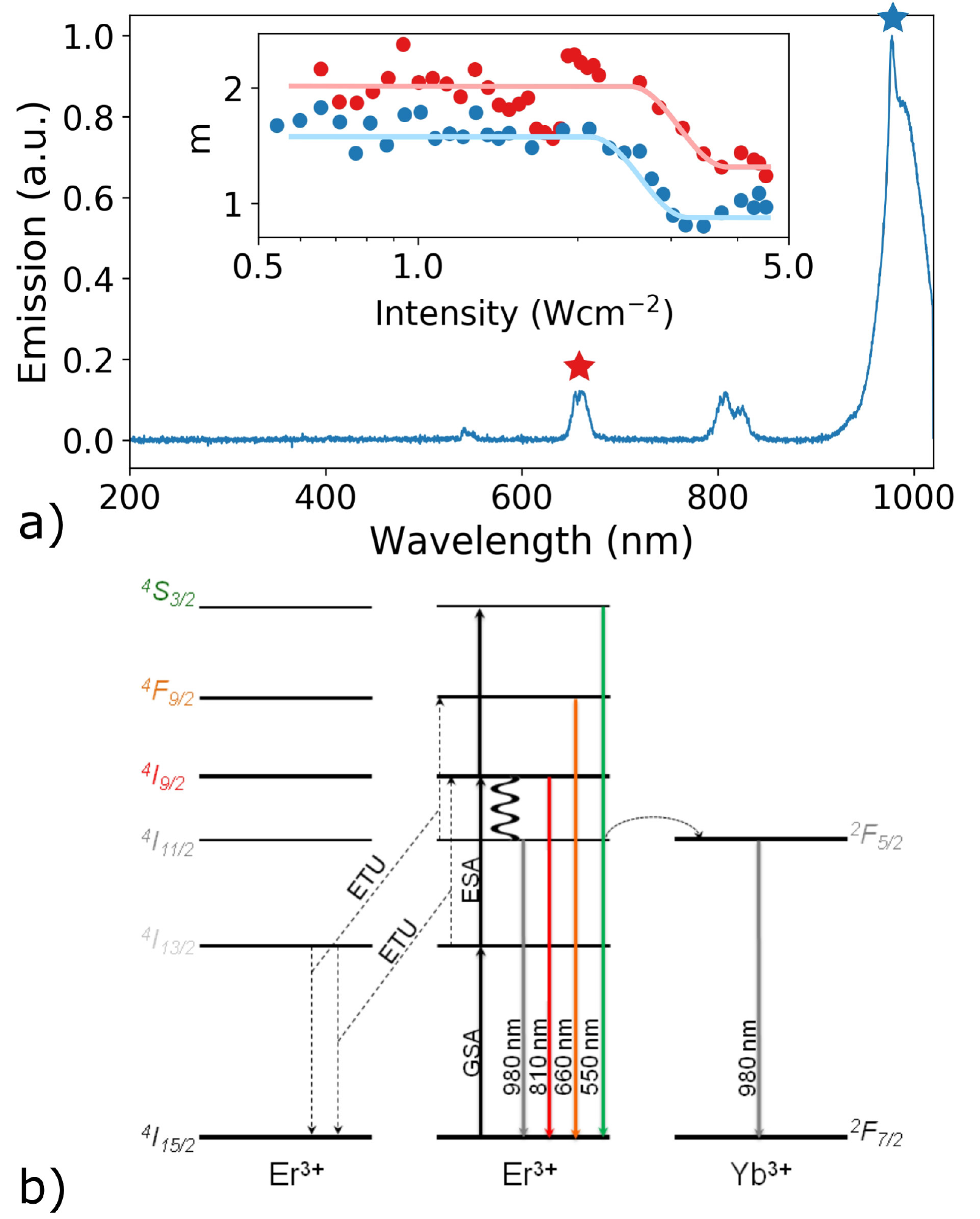}
    \caption{a) Fluorescence of embedded lanthanide-doped particles under 1552\nm\ excitation. The inset shows the local exponent of the emission intensity ($m$ in equation \ref{eq:linear_fit}) as a function of the excitation intensity, plotted versus the excitation intensity for the transitions at 660\nm\ and 980\nm\ wavelength (marked with stars in main figure). In each case a guide to the eye is included to show the transition between low intensity and saturation regimes. (b) The energy levels of Er\textsuperscript{3+} ion leading to the measured fluorescence are shown. The ground state absorption (GSA) occurs pumping level \textsuperscript{4}I\textsubscript{$\frac{15}{2}$} , from where excited state absorption (ESA) to higher energy levels occurs. Solid and dashed lines display radiative and energy transfer (ETU) respectively, while curvy lines show multiphonon relaxation.}
    \label{fig:FlourescenceResults}
\end{figure*}

The fluorescence spectrum corresponding to an excitation intensity of $2.8\;\textrm{Wcm}^{-2}$ at $1552\;\textrm{nm}$ wavelength is shown in figure $\ref{fig:FlourescenceResults}$(a). The spectrum shows several emission peaks in the visible and the near infrared despite the much longer excitation wavelength. The wavelengths of these peaks are in good agreement with those predicted by the various up-conversion pathways present in the energy level diagram shown in (b). Emission around 980\nm, 810\nm, 660\nm\ and 550\nm\ can be seen when exciting at 1552\nm. The strongest emission is at 980\nm. Emission at 810\nm\ occurs from the \textsuperscript{4}I\textsubscript{$\frac{9}{2}$} erbium level. Population of this level can occur by up-conversion of two 1552\nm\ photons either by excited state absorption (ESA) or energy transfer up-conversion (ETU) to a nearby erbium ion. The concentration of erbium was 3\%, suggesting up-conversion through ESA is more probable than ETU. The energy spacing between the \textsuperscript{4}I\textsubscript{$\frac{9}{2}$} and \textsuperscript{4}I\textsubscript{$\frac{11}{2}$} levels in NaYF\textsubscript{4} favours multiphonon relaxation with further emission at 980\nm. A concentration of Yb at 18\%, makes energy transfer to the ion probable, with further emissions at 980\nm. Emission at 660\nm\ is probable via ETU to the \textsuperscript{4}F\textsubscript{$\frac{9}{2}$} erbium level, while 550\nm\ emission requires a population of the \textsuperscript{4}S\textsubscript{$\frac{3}{2}$} level by three photon up-conversion through sequential absorptions.



The exponent of the fluorescence signal as a function of the excitation intensity, can be used to characterize the non-linearity of a process. Taking the logarithm of the signal and excitation intensity allows the local exponent to be determined via the local gradient ($m$ in equation \ref{eq:linear_fit}). The inset of Figure $\ref{fig:FlourescenceResults}$(a) shows the excitation intensity dependence of the emission intensity gradient for the \textsuperscript{4}F\textsubscript{$\frac{9}{2}$}$\rightarrow$\textsuperscript{4}I\textsubscript{$\frac{15}{2}$} and \textsuperscript{4}I\textsubscript{$\frac{11}{2}$}$\rightarrow$\textsuperscript{4}I\textsubscript{$\frac{15}{2}$}  transitions in Er\textsuperscript{3+} ions. The emission wavelengths of these transitions are 660\nm\ and 980\nm, respectively. Although the emission signal is lower at 660\nm\ compared to at 980\nm, the transition avoids quenching through absorption by Yb ions. This simplifies the analysis of the intensity dependence. The values in the inset show the local gradient of the emission intensity, calculated using a forward difference scheme and averaged over four neighbours. This allows for the transition to the saturation regime to be quantified, such as in \cite{Wuerth2020}.

Two regions can be identified depending on the intensity of the excitation source. In the low-intensity regime, without any loss mechanisms, $m$ would describe how many photons from the excitation are absorbed in the UC process. From the energy level diagram in figure \ref{fig:FlourescenceResults}(b) we can discern that the emission at 660\nm\ and 980\nm\ are two photon processes, meaning that the gradient could possibly reach two for low excitation intensities. However, due to loss mechanisms such as thermal effects, and pump profile mismatch, the experimental value of $m$ may be below the number of photons in the process \cite{Pollnau2000}. The gradient remains at 2.0 and 1.7 for the 660\nm\ and 980\nm\ transition, respectively, over most of the range of incident intensities. As the incident intensity increases, above 2.6 Wcm\textsuperscript{-2} in our experimental setup, the local gradient drops to below 1.5 and 1.0 for the 660\nm\ and 980\nm\ transition, respectively. This indicates the transition to the saturation regime. In the saturation regime, relaxation via the up-conversion pathway becomes dominant, causing a linear relationship between incident photons and the up-conversion process.

\begin{figure*}[h!]
    \centering
    \includegraphics[width=1.0\textwidth]{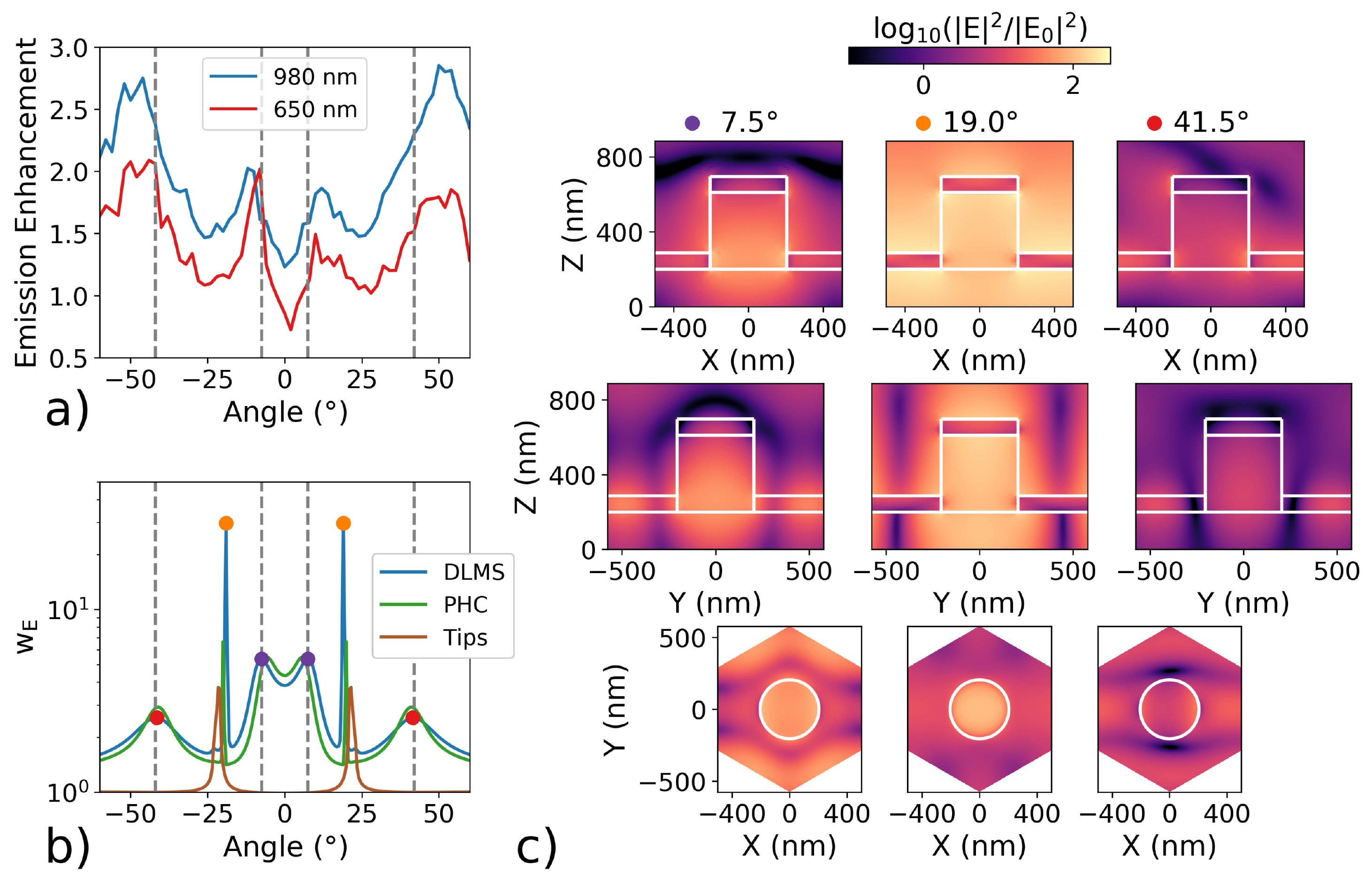}
    \caption{(a) The measured enhancement comparing emission from embedded up-conversion particles on a double-layer metasurface compared to planar silicon. The enhancement is shown for an emission wavelength of 660 and 980\nm. Excitation is at 1552\nm. (b) The simulated electric field energy enhancement in close to the coated double-layer metasurface. Three peaks are marked with purple, orange and red dots. (c) Logarithm of the squared electric field enhancement associated with the resonance peaks from (b). Each column represents a different resonance angle, while each row represents a different cross-sectional slice of the simulation.}
    \label{fig:EnhancementAnalysis}
\end{figure*}

The experimental emission enhancement as a function of incident angle for embedded particles on a double-layer metasurface compared to on planar silicon is shown in figure \ref{fig:EnhancementAnalysis}(a). The emission enhancement is shown for emission at both 980\nm\ and 660\nm. From the experimentally determined enhancement, two sets of angular symmetric peaks at incident angles of $\pm 10\degree$ and $\pm 50\degree$ are visible. The enhancement peak at $\pm 50\degree$ is around 50\% higher than for the peak at $\pm 10\degree$. This trend is visible for the emission at both 980 and 660\nm\ emission. The emission enhancement by using the double-layer metasurface instead of a planar silicon substrate amounts to around 2.7 and 2.0 for the transitions with 980\nm\ and 660\nm\ emission, respectively. We do not expect much higher values as only a small part of the excited up-conversion particles are affected by the metasurface. It should be noted that in order to achieve the highest signal to noise ratio, the measurements of the emission enhancement were performed using laser intensities in the saturation region (see figure \ref{fig:FlourescenceResults}), which limits the measurable value of the enhancement factor. 

In contrast, the simulated the electric field energy close to the double-layer metasurface, figure \ref{fig:EnhancementAnalysis}(b), shows a small angle peak at $\pm 7.5\degree$ that is more pronounced than the high angle peak at $\pm 41.5\degree$. A further extremely sharp peak at $\pm 19\degree$ is also visible. Also shown is the electric field energy in the near field for the photonic crystal slab and silicon tip structures (shown schematically in figure \ref{fig:GeometryVariation}). The small and large angle peaks are clearly due to modes associated with the PhC slab. Interestingly, the sharp resonance at $19\degree$ seems to be a combination of both a PhC and tip mode, with an electric field energy higher than that provided by either structure individually. This suggests that the electric near field energy of a photonic crystal slab can be enhanced by moving to the double-layer metasurface presented here.

The sharp peak in electric field energy at $19\degree$ is due to modes associated with the disappearance of a diffraction order, i.e. the Rayleigh-Wood anomaly discussed previously. These modes are not present in the experimental enhancement measurements, in the same way that the resonances in transmission measurements were also not visible in figure \ref{fig:ARDT}. This can be attributed to the extremely fine bandwidth of the resonances, making their measurement difficult with finite line-width sources. Moreover, the resonances themselves are extremely sensitive with regards to inhomogeneities in the metasurface. Any imperfections of the periodic lattice are likely to provide strong damping for these resonances.

The angular position of both the small and large angle resonance are slightly underestimated by the simulation when compared to the experiment. Figure \ref{fig:ARDT} shows that both of these modes have a positive dispersion in this wavelength region, i.e. they have increasing resonance angle for decreasing wavelength. It has been previously established that the wavelength of resonance modes can be shifted via straightforward geometrical manipulations, such as varying the thickness of a photonic crystal slab \cite{Ahiboz2020}. If the metasurface resonance modes were shifted to a slightly longer wavelength, the resonances would appear at higher angles at the excitation wavelength of 1552\nm. Thus, the discrepancy in angular positions of the resonances when comparing simulation and experiment is likely due to a small discrepancy in the geometry of the metasurface between simulation and experiment.


The discrepancy in the small and large angle enhancement factors between experiment and simulation might depend on the local distribution of the electric field close to the surface. If high field intensities are localized very close to the surface, they may be difficult for larger particles to access. To investigate this, the near field distribution in the \textit{x-z}, \textit{y-z} and \textit{x-y} planes, as shown in figure \ref{fig:EnhancementAnalysis}(c), were calculated. The modes which provided an experimentally observed emission enhancement were those which had a peak in the simulation at 7.5 and 41.5$\degree$. These modes appear to be mainly localized to inside the lower silicon layer of the double-layer metasurface. The enhanced near fields inside the coating layer at each of these two resonances do not appear to differ strongly in terms of how localized the fields are. Therefore this is unlikely to account of the difference in experimentally observed emission enhancement. When looking at the localized fields for the resonance at 19$\degree$ in the \textit{y-z} plane, the pattern of the field appears reminiscent of a plane wave excitation travelling in the \textit{x-y} plane. This further supports the idea that this resonance can be attributed to coupling to Bloch wave, i.e. a Rayleigh-Wood anomaly. 

In order to account for the discrepancy between the experimentally measured emission enhancement and the simulated electric field energy enhancement we move from near field effects to consider the far field, specifically the effect of light trapping. The light reflected for $7.5\degree$ incidence is completely specular, i.e. it is confined to the zeroth diffraction order. Due to this, the reflected light can escape from the coating layer. For light incident at $41.5\degree$ the reflected light is only 15\% specular. The rest of the reflection is divided equally between two first order diffraction orders with angles of $59.6\degree$ to the surface normal. This is well over the critical angle of $43.1\degree$ for light to escape the coating layer at 1552\nm\ wavelength. This means the light is trapped inside the coating, with a certain finite lifetime due to both scattering and absorption from the particles and further interactions with the double-layer metasurface. 

This supports the previous result that the two resonances which contributed to the experimentally observed emission enhancement at 1552\nm\ excitation wavelength were mainly due to the photonic crystal slab. However, the addition of the silicon tips does work to broaden these resonances, at least for small incident angles. This could be useful for increasing the bandwidth of these resonant features. Furthermore, the presence of the tips causes the Rayleigh-Wood anomalies of the double-layer metasurface to split into two distinct sets of resonances, neither of which are at precisely the same spectral positions as the resonances found in their constituent parts (the tips or PhC in isolation). With correct engineering of the tip geometry, this could be used to overlap the Rayleigh-Wood anomalies with the leaky eigenmode resonances of the PhC slab structure to enhance the light interaction at certain spectral positions.

\begin{figure*}[h!]
    \centering
    \includegraphics[width=1.0\textwidth]{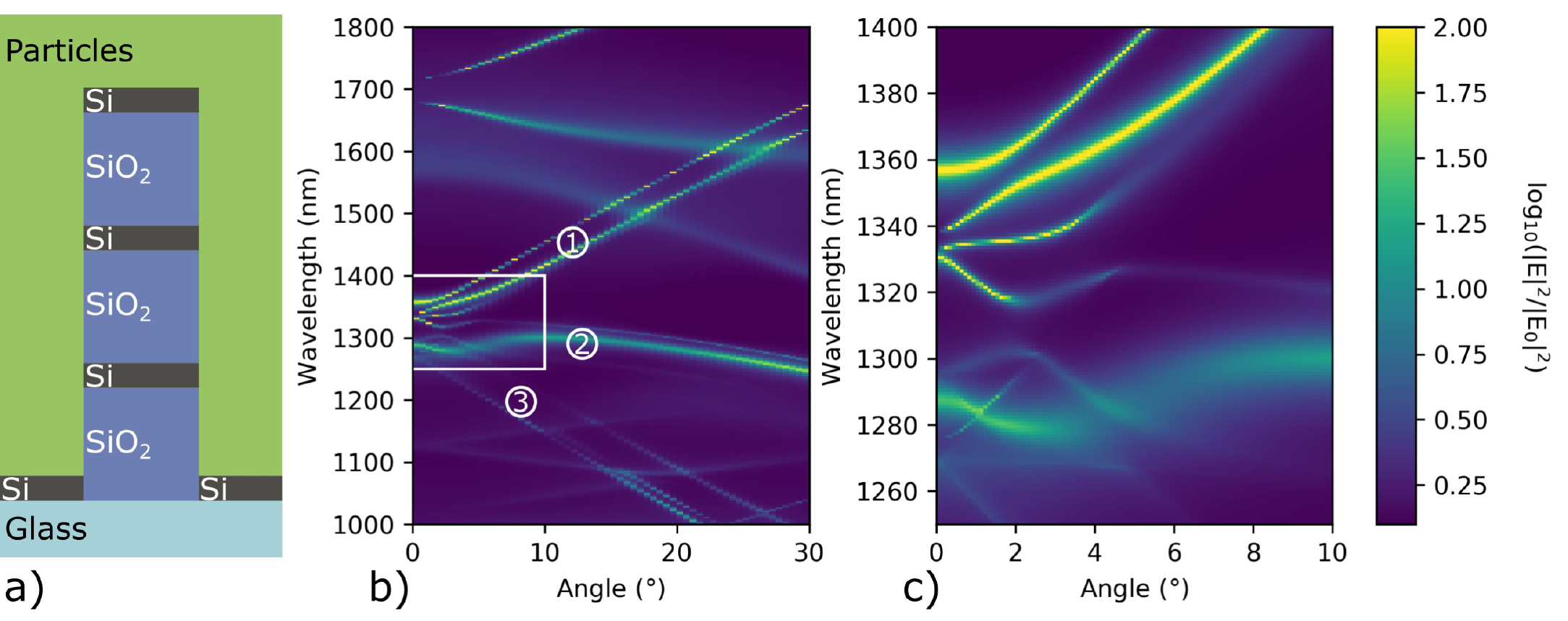}
    \caption{(a) schematic of the cross section of the 4-layer metasurface. (b) and (c) The logarithm of the electric field intensity enhancement in the coating region directly above the metasurface (green region in (a)). The three sets of split Rayleigh-Wood anomalies are labeled 1-3. (c) shows the data for the rectangular region shown in (b) but for higher resolution.}
    \label{fig:Multilayer}
\end{figure*}

In order to further investigate the ability of such multi-layer metasurfaces to split the Rayleigh-Wood anomalies into multiple distinct resonances, a four-layer metasurface was simulated. Figure \ref{fig:Multilayer}(a) shows a schematic image of the structure. The base layer remains the photonic crystal slab of hexagonally arranged holes in a silicon layer. Inside the holes of the photonic crystal slab there is a multi-layer nanorod consisting of three repetitions of a SiO$_{2}$ nanorod with a thinner silicon nanodisc. The end result is four layers of high index material (silicon) in a periodic arrangement. The electric field energy enhancement surrounding the nanorods and for 200\nm\ above the surface is shown in figure \ref{fig:Multilayer}(b). The majority of the resonances, those that were not classified as Rayleigh-Wood anomalies, retain their spectral positions and relative intensities. The Rayleigh-Wood anomalies split into a groups of three resonances. In figure \ref{fig:GeometryVariation}(b,c) it was demonstrated that the Rayleigh-Wood anomalies for single layer metasurfaces are present in three branches. These three branches are due to the combination of the hexagonal symmetry of the lattice and the incident angle varying along a symmetry direction ($\Gamma-\mathrm{K}$). Each branch is then further subdivided due to the presence of multiple layers in the nanorod part of the metasurface. This can be seen in figure \ref{fig:Multilayer}(b) which presents three sets of resonance groups with very different dispersions (as in \ref{fig:GeometryVariation}(b,c)). The groups are labeled 1-3 and each of these groups contains three resonances with a similar dispersion. Figure \ref{fig:Multilayer}(c) shows the behavior close to normal incidence, revealing up to nine resonances in this spectral region. This large number of high quality resonances is able to cover much of the wavelength spectrum for small incident angles. Multi-layer metasurfaces as presented here may offer the ability to create broadband strong light-matter interactions via their enhanced near fields. This would make multi-layer metasurfaces a key enabler for applications such as up-conversion of light for solar energy applications.

\FloatBarrier

\section{Conclusion}

We presented a stacked, double-layer metasurface consisting of a traditional hexagonal lattice photonic crystal slab and integrated SiO\textsubscript{2} nanopillars with silicon nanodisc tips. The metasurface was subsequently coated with lanthanide-doped \UCParticle\  particles  dispersed in PDMS. We performed angular resolved direct transmission measurements on the coated and uncoated metasurface. These measurements revealed a rich pattern of resonances in the transmission. By performing finite element simulations, we showed that the transmission resonances are accompanied with strongly enhanced near fields close to the metasurface. A subset of the modes were shown to be Rayleigh-Wood anomaly type resonances associated with the disappearance of diffraction orders. These modes presented the strongest enhancement in the optical simulations, which was attributed to the interaction of the two layers of the metasurface. We found that the major difference in the double-layer metasurface when compared to a conventional single-layer metasurface appears to be in these Rayleigh-Wood anomalies, which are split into two different sets of modes and spectrally broadened. 

Spectral fluorescence measurements upon excitation at 1552\nm\ revealed emission bands occurring at the wavelengths associated with the up-conversion process in lanthanide particles. 
By varying the angle of incidence we could show an increase in emission intensity for embedded particles on the double metasurface of up to 2.7 compared to on a planar silicon surface, despite only around 1\% of the up-conversion material being exposed to enhanced near-fields. The angles of peak enhancement were in good agreement with those predicted from simulation, meaning that we could attribute the enhanced emission to the leaky mode resonances of the double-layer metasurface. The enhancement was attributed to the combined effect of enhanced near fields and light trapping due to light scattered by the resonant modes. The nature of the resonances was further investigated, confirming that the experimentally observed resonances were mainly due to the presence of the lower photonic crystal slab part of the metasurface. The role of the tips was to broaden the preexisting resonances and provide a splitting of the Rayleigh-Wood anomaly type resonances. A final simulation reveals that stacking of multiple meatsurface layers yields splitting into multiple different modes. Since these resonances are particularly intense, this could be used to engineer their position and number for applications that require multiple sharp resonances, e.g. due to a broad excitation spectrum.

These results make clear that mutli-layer metasurfaces can be used to benefit the photon up-conversion upon low-intensity and broadband excitation. This enhancement could be further improved by decreasing the extent of the up-conversion nanoparticles, so that their volume overlaps with the mode volume of the leaky resonance modes of the metasurface. The improved up-conversion of infrared light shown here will enable new possibilites for the spectral manipulation of light, including expanding the exploitable range of wavelengths for solar cell applications.

\section*{Funding Information}
Partially funded by the Deutsche Forschungsgemeinschaft (DFG, German Research Foundation) under Germany's Excellence Strategy – The Berlin Mathematics Research Center MATH+ (EXC-2046/1, project ID: 390685689, AA4-6) as well as the Helmholtz Association for funding the Helmholtz Excellence Network SOLARMATH, a strategic collaboration of Helmholtz-Zentrum Berlin and MATH+ (grant no. ExNet-0042-Phase-2-3). The simulations were done at the Berlin Joint Lab for Optical Simulations for Energy Research (BerOSE).

\section*{Acknowledgements}

The authors thank Lin Zschiedrich for fruitful discussion on optical simulations, Carola Klimm for obtaining SEM images and Daniel Amkreutz for electron beam evaporation of silicon.

\section*{Disclosures}
The authors declare that there are no conflicts of interest related to this article.


\bibliography{LED2020}
\bibliographystyle{unsrt}

\end{document}